\pdfoutput=1
\documentclass[11pt]{article}

\usepackage{jheppub}
\usepackage{amsmath}
\usepackage{verbatim}
\usepackage{amssymb}
\usepackage{inputenc}
\usepackage{textcomp}
\usepackage{appendix}
\usepackage{mathtools}
\usepackage{multirow}
\usepackage{mathrsfs}
\usepackage{stmaryrd}
\usepackage{blindtext}

\newcommand{\e}{\epsilon}
\newcommand{\be}[1]{\begin{equation}\label{#1} }
\newcommand{\ee}{\end{equation}}
\newcommand{\bea}[1]{\begin{eqnarray}\label{#1} }
\newcommand{\eea}{\end{eqnarray}}
\newcommand{\bes}{\begin{subequations}}
\newcommand{\ees}{\end{subequations}}

\newcommand{\p}{\partial}

\newcommand{\non}{\nonumber}

\newcommand{\D}{\Delta}

\renewcommand{\a}{\alpha}

\renewcommand{\t}{\tau}
\newcommand{\s}{\sigma}

\newcommand{\pr}{\prime}
\usepackage{pifont}

\title{Circuit Complexity for Carrollian  Conformal (BMS) Field Theories}

\author{Arpan Bhattacharyya,} \author{Poulami Nandi\footnote{corresponding author.}.}  \author{\\}

\affiliation{Indian Institute of Technology Gandhinagar, Palaj 382355 Gujarat. India. \\}
 
\emailAdd{abhattacharyya@iitgn.ac.in, poulami.n@iitgn.ac.in}

\abstract{We systematically explore the construction of Nielsen's circuit complexity to a  non-Lorentzian field theory keeping in mind its connection with flat holography. We consider a $2d$ boundary field theory dual to $3d$ asymptotically flat spacetimes with infinite-dimensional BMS$_3$  as the asymptotic symmetry algebra. We compute the circuit complexity functional in two distinct ways. For the Virasoro group, the complexity functional resembles the geometric action on its co-adjoint orbit. Using the limiting approach on the relativistic results, we show that it is possible to write BMS complexity in terms of the geometric action on BMS co-adjoint orbit. However, the limiting approach fails to capture essential information about the conserved currents generating BMS supertranslations. Hence, we refine our analysis using the intrinsic approach. Here, we use only the symmetry transformations and group product laws of BMS to write the complexity functional. The refined analysis shows a richer structure than only the geometric action. Lastly, we extremize and solve the equations of motion (for a simple solution) in terms of the group paths and connect our results with available literature. } 

\begin{document}

\maketitle
\section{Introduction}
The entropy of the black hole, containing the information of its microstates, is proportional to the area of its event horizon and not its volume. With this simple yet, counter-intuitive statement, we witnessed the first notion of the Holographic Principle \cite{tHooft:1993dmi,Susskind:1994vu}. The holographic principle says that we can study a theory of quantum gravity by looking into a field theory without gravity, living on its boundary in one lower dimension. The most powerful and successful example of the Holographic Principle was given by the AdS/CFT correspondence \cite{Maldacena:1997re,Gubser:1998bc,Witten:1998qj}. To date, it has found many interesting applications in various branches of physics \cite{Kovtun:2003wp,McGreevy:2009xe,Hartnoll:2009sz,Harlow:2014yka,Ammon:2015wua,2017LNP...931.....R,2022EPJC...82..458B}. Even after the enormous success, our physical world can not be described by a constant negative curvature spacetime. For many astrophysical purposes, it is best approximated as asymptotically flat instead of AdS. 

\smallskip
\noindent
There has been quite some advancement in the program of flat space holography over the past few years \cite{Bagchi:2010zz,Barnich:2006av,Barnich:2010eb}. It is done mainly by looking into the symmetries of the bulk asymptotically flat spacetimes and proposing the boundary field theory to inherit them. The symmetries at the boundary of $3d$ and $4d$ asymptotically flat spacetimes are enhanced from the Poincar\'e symmetries and are given by the infinite-dimensional Bondi-Metzner-Sachs (BMS) group \cite{Bondi:1962px,Sachs:1962zza}. Hence, the dual boundary field theories are also expected to be BMS invariant. The initial attempts to build the holographic principle for flat spacetimes consider taking the AdS radius to infinity, which in turn introduces an ultra-relativistic  (Carrollian \cite{Leblond65,Sengupta65})  limit on the dual boundary Conformal Field Theory \cite{Bagchi:2012cy}. This line of approach is widely known as Carrollian holography.\footnote{There exists another parallel approach to flat space holography, called the Celestial holography. For $4d$ asymptotically flat spacetimes, the structure at null infinity $\mathscr{I}^\pm$ is $\mathbb{R}\times \mathbb{S}^2$. On the sphere at infinity, called the celestial sphere, the structure is
very much like $2d$ CFT. Hence, the physics of $4d$ flat spacetimes can be expressed in terms of $2d$ CFT, giving rise to a correspondence between the asymptotic symmetries and scattering amplitudes. For more details, the readers are referred to the excellent reviews \cite{Strominger:2017zoo,Pasterski:2021rjz,Raclariu:2021zjz}, and also the recent works on bridging these two considerably different flat holography pictures, Carrollian and Celestial \cite{Donnay:2022aba,Bagchi:2022emh}.} It was also shown that the BMS algebra is isomorphic to the Carrollian version of Conformal Algebra, or Carrollian Conformal Algebra (CCA). The field theories having Carrollian Conformal symmetries live on the null boundaries ($\mathscr{I}^\pm$) of the asymptotically flat spacetimes. In the Carrollian limit, the lightcone collapses, and the speed of light goes to zero. On the null boundary, the metric becomes degenerate, and the underlying geometry is no longer (pseudo) Riemannian. The ultra-relativistic limit replaces the relativistic Poincare symmetry to Carrollian symmetries. However, it is well known now that Carrollian symmetries are associated with any null manifold, e.g. the event horizon of the black hole \cite{Donnay:2019jiz}. Thus, Carrollian and Conformal Carrollian symmetries are vividly related to flat space holography and black hole physics. There has been significant success to this formulation, specifically in $3d$ bulk and $2d$ boundary \cite{Bagchi:2012yk}, BMS Cardy type formula \cite{Bagchi:2012xr,Barnich:2012xq}, entanglement entropy \cite{Bagchi:2014iea,Jiang:2017ecm,Hijano:2017eii}, holographic construction of correlation functions \cite{Bagchi:2015wna} etc., to name a few. Apart from its importance in the flat holography program, Carrollian physics has found important applications in condensed matter physics through the theory of fractons \cite{Bidussi:2021nmp}, flat bands \cite{Bagchi:2022eui}, cosmology, dark energy and inflation \cite{deBoer:2021jej}, Carroll fluids \cite{deBoer:2017ing,Ciambelli:2018wre}, the tensionless limit of string theory \cite{Bagchi:2013bga,Bagchi:2019cay} and Carrollian gravity \cite{Bergshoeff:2017btm,Duval:2017els}.

\smallskip
\noindent
On the other hand, using quantum information theory, one aims to understand the entanglement between two or more subsystems. The entanglement entropy (EE) is one of the useful measurements of the entanglement between two subsystems, which can be computed by tracing out one of the subsystems and computing the Von Neumann entropy for the resulting reduced density matrix. The relation between quantum information theory and the holographic principle came into prominence with the formulation of the Ryu-Takayanagi (RT) proposal \cite{Ryu:2006bv,Ryu:2006ef} (later, a covariant version of this proposal had been put forward in  \cite{Hubeny:2007xt}) of entanglement entropy using AdS/CFT. It says that the quantum correlations between two regions in a CFT are given by the areas of the minimal surfaces in asymptotically AdS spacetimes. The geometrization of EE has been insightful in studying the encoding of boundary information in the bulk and vice versa. 
Even though it partially addresses the long-standing questions in theoretical physics, e.g. the famous information loss paradox \cite{Penington:2019npb,Almheiri:2019psf,Penington:2019kki,Almheiri:2019qdq}\footnote{For an interesting review, please see \cite{Raju:2020smc} and the references therein.}, it is now understood that EE alone is not sufficient to capture the entire information on quantum correlations in a state \cite{Susskind:2014moa}. 

\smallskip
\noindent
In the last few years, another measurement of quantum information rose into eminence, called the quantum computational complexity \cite{2006Sci...311.1133N,nielsen2005geometric}. In simplistic notion, quantum computational complexity determines how difficult it is to reach from a certain reference state to a target state using unitary operations (gates) \cite{nielsen2005geometric,watrous2008quantum,aaronson2016complexity,2006Sci...311.1133N}. It essentially describes an extremal circuit consisting of quantum gates starting from the reference to the target state for discrete systems or a geodesic distance in the manifold of unitary (group) operators for field theories . In the context of AdS/CFT, the growth of holographic complexity can capture the growth of volume behind blackhole interiors. The two most notable holographic complexity proposals are known as CV (complexity=volume) \cite{Stanford:2014jda} and CA (complexity=action) \cite{Brown:2015bva}. These two geometric quantities have been conjectured to be dual to the circuit complexity of the field theory state. After that, there has been an upsurge in understanding complexity in the context of field theory \cite{Jefferson:2017sdb,Chapman:2017rqy, Hackl:2018ptj,Khan:2018rzm,Bhattacharyya:2018bbv,Bhattacharyya:2019kvj,me1,Magan:2018nmu,Caputa:2018kdj,Erdmenger:2020sup,Flory:2020eot,Flory:2020dja,Erdmenger:2021wzc,Chagnet:2021uvi,Koch:2021tvp,Bhattacharyya:2022ren,2022PhRvD.106h6010B,Erdmenger:2022lov,Rabambi:2022jwu,Jiang:2018nzg,Guo:2020dsi} \footnote{This list is by no means exhaustive. Interested readers are referred to these reviews \cite{Chapman:2021jbh,Bhattacharyya:2021cwf}, and citations therein for more details.}. 

\smallskip
\noindent
Despite the rise of quantum complexity in theoretical physics, it is mostly focused on applications in AdS/CFT and quantum many-body systems. Hence, it is an opportune moment to systematically explore the notion of the circuit  complexity to other non-Lorentzian field theories and flat space holography. For earlier work in this direction, see \cite{Fareghbal:2018ngr,Banerjee:2022ime}. These works mainly focus either on the holographic side or explore circuit complexity for some simple quantum mechanical models. So far, a study of circuit complexity, purely from the field theory for BMS field theories, has yet to be done. With this aim, we take the first step towards this direction in our current work. Also, we are considering a semi-direct product group instead of the usual direct product (e.g., centrally extended Virasoro and Kac-Moody) groups to compute the circuit complexity.

\paragraph{Objectives of the paper:} In this paper, we systematically explore the construction of Nielsen's circuit complexity for a  non-Lorentzian field theory keeping in mind its connection with flat holography. We consider a $2d$ boundary field theory dual to $3d$ asymptotically flat spacetimes with BMS$_3$ as the asymptotic symmetry algebra. We aim to keep this paper as self-contained as possible. In section \ref{bmsrevisit}, we briefly review the Bondi-Metzner-Sachs (BMS$_3$) group, its algebra, representation theory, and also the contraction from  $2d$ relativistic Conformal algebra. We discuss the isomorphism of BMS$_3$ algebra with the Carrollian Conformal Algebra (CCA). It is possible to reach CCA as an ultra-relativistic ($c\to 0$) limit\footnote{The ultra-relativistic limit of the relativistic CFT algebra gives only CCA of a particular level $N=2$. The isomorphism between BMS and CCA is also for the level $N=2$. In this paper, we only focus on CCA$^{(N=2)}$. For more details, the readers are referred to \cite{Duval:2014uva,Bagchi:2019clu}.} to the relativistic CFT algebra. The field theories having  Carrollian Conformal symmetries (Carrollian CFTs) are putative dual to asymptotically flat spacetimes. We also mention that for $2d$ boundary, there exists another isomorphism between BMS$_3$ and the Galilean Conformal Algebra (GCA), which is the non-relativistic limit ($c\to \infty$) of  CFT algebra. We use this isomorphism in the later section. Next, in section \ref{vircom}, we revisit the construction of circuit complexity for  $2d$ CFTs, the infinite-dimensional Virasoro algebra. We discuss that Nielsen's complexity functional (with essential modifications for an infinite dimensional system) for the Virasoro group can be given by the geometric action on the co-adjoint orbit of the Virasoro group. We review the construction of the gates and instantaneous velocity $\e(\tau,\s)$ from group elements product law and draw a parallel with the Maurer-Cartan form before adding the extra contribution from the central extension of the Virasoro group. Then, we move towards the main objective of this paper, i.e. the circuit complexity calculation for $2d$ BMS$_3$ invariant field theories (or, equivalently, $2d$ Carrollian CFTs). First, in section \ref{liman}, we follow the limiting analysis on the relativistic CFT results to reach the complexity functional of $2d$ Carrollian CFT. We find that following the non-relativistic contraction of the CFT complexity functional, it is possible to obtain the BMS complexity functional, resembling the BMS geometric action. However, this analysis does not include all of the information about the BMS conserved currents (namely, the supertranslations), and hence we refine our analysis in section \ref{intan}. We complete this section purely from the symmetries of BMS$_3$ and construct the complexity functional intrinsically. We see that the BMS complexity functional has a richer structure and can be reduced to the geometric action on the co-adjoint orbit only for specific values of the instantaneous velocity. Later, we extremize the complexity functional and find the optimal path (for the simplest solution) in the manifold of unitary transformations to reach the target state from the chosen reference state (BMS primaries). We comment on our results and their similarities with available literature in parallel with the Virasoro circuit complexity. We end this paper with a discussion and future extensions in section \ref{discussion}.

\section{Carrollian Conformal (BMS$_3$) physics: prerequisites}\label{bmsrevisit}
Let us start with the necessary backgrounds on the $2d$ BMS$_3$ invariant field theories (Carrollian CFTs). This will set the stage for the main objectives of the paper. The asymptotic structure at the null boundary $\mathscr{I}^{\pm}$ of $3d$ asymptotically flat spacetimes are given by $\mathbb{R}_u\times \mathbb{S}^1$. Here, the null direction is $\mathbb{R}_u$ with  $u,v=t\pm x$, and $\mathbb{S}^1$ gives the circle at infinity. The symmetries at the boundary of the spacetimes are given by the asymptotic symmetry groups (ASG), and the associated algebra is called the asymptotic symmetry algebra (ASA). It was shown by Bondi, van der Burgh, Metzner, and Sachs that the ASG for $3d$ asymptotically flat spacetimes are given by the infinite-dimensional Bondi-Metzner-Sachs (BMS) group \cite{Bondi:1962px,Sachs:1962zza}, which is different from the usual Poincar\'e group. BMS group can be realised as the semi-direct product of all local conformal transformations (diffeomorphisms),  and the angle-dependent translations of the (null) time direction on the circle $\mathbb{S}^1$ called the supertranslations. In this section, we review some useful features of the $3d$ BMS group.

\subsection{Group algebra and contraction}
The BMS$_3$ algebra is given by,
\bes{}\label{bms3al}
\begin{eqnarray}
\left[L_m,L_n\right]&=&(m-n)L_{m+n}+\frac{c_L}{12}(n^3-n) \delta_{m+n,0},\\
\left[L_m,M_n\right]&=&(m-n)M_{m+n}+\frac{c_M}{12}(n^3-n) \delta_{m+n,0},\\
\left[M_n,M_m\right]&=&0.
\end{eqnarray}
\ees
Here, $L_n$'s, which define the diffeomorphisms on the circle $\mathbb{S}^1$, are the generators of the superrotations. On the other hand, $M_n$'s are the generators of the supertranslations or the angle-dependent translations on the circle at infinity. $c_L$ and $c_M$ are the two central extensions. For the Einstein gravity, they turn out to be the following $$c_L=0\quad \textrm{and}\quad c_M=\frac{3}{G},$$ where $G$ is Newton's Constant \cite{Barnich:2006av}. This is obtained by performing  an asymptotic symmetry analysis and computing the Poisson bracket for the charges.

\medskip
\noindent
 In a seminal analysis by Brown and Henneaux \cite{Brown:1986nw}, the ASA for $3d$ Anti-de Sitter spacetime was also found to be infinite-dimensional, comprising of two copies of the Virasoro algebra. 
 \bes{}
\label{viraal}
\begin{eqnarray}
\left[\mathcal{L}_m,\mathcal{L}_n\right]&=&(m-n)\mathcal{L}_{m+n}+\frac{c}{12}(n^3-n) \delta_{m+n,0},\\
\left[\bar{\mathcal{L}}_m,\bar{\mathcal{L}}_n\right]&=&(m-n)\bar{\mathcal{L}}_{m+n}+\frac{\bar{c}}{12}(n^3-n) \delta_{m+n,0}.
\end{eqnarray}
\ees
Here, the central terms for Einstein gravity in AdS$_3$ are given by, 
\be{}
c=\bar{c}=\frac{3l}{2G}.
\ee
with $l$ being the AdS radius. The BMS$_3$ algebra \eqref{bms3al} of asymptotically flat spacetime can also be reached from the ASA of the AdS$_3$ \eqref{viraal} by an In\"on\"u-Wigner contraction,
\be{urbmsgen}
L_n=\mathcal{L}_n-\bar{\mathcal{L}}_{-n},~~~ M_n=\epsilon(\mathcal{L}_n+\bar{\mathcal{L}}_{-n}),~~\e \to 0.
\ee
In literature, this is referred to as the ultra-relativistic or Carrollian limit \cite{Leblond65,Sengupta65}. In the Carrollian limit, the light cone closes up, and the speed of light goes to zero ($c\to 0$). This can be understood by the following spacetime contraction
\be{}
t\to \e t,~~ x^i \to x^i, ~~ \e\to 0. \implies \frac{v}{c}=\frac{1}{c}\frac{x}{ t}\to \infty. 
\ee
The Carrollian limit can be interpreted as taking the AdS radius $(l)$ to infinity in bulk, or equivalently taking the inverse radius $\e=\frac{1}{l}\to 0$ \cite{Bagchi:2012cy}. Also, we can recover the BMS central terms from the Brown-Henneaux central charges of AdS$_3$ using the following contraction,
\be{}
c_L=(c-\bar{c})=0, ~~~ c_M=\lim_{l \to \infty}\frac{1}{l}(c+\bar{c})=\frac{3}{G}.
\ee

\bigskip
\noindent
On the other hand, the Virasoro generators on the cylinder (corresponding to the conformal boundary of $3d$ global AdS) 
  \be{}
 \mathcal{L}_m=e^{im \omega}\p_{\omega},~~\bar{\mathcal{L}}_m=e^{-im \bar{\omega}}\p_{\bar{\omega}}, ~~ \text{where,~}\omega,\bar{\omega}=t \pm \sigma
 \ee
 under the In\"on\"u-Wigner contraction
\be{}
\s \to \s,~~ t \to\e t, ~~\e \to 0, \implies\frac{v}{c}\sim \frac{1}{c}\frac{\s}{t}\to \infty
\ee 
gives the BMS$_3$ generators on the null cylinder $(\s,t)$ 
 \be{}
 L_n= e^{in \s}(\p_\s+in t \p_t),~~ M_n=i e^{in \s}\p_t.
 \ee
This is known as the cylinder representation of BMS, where only the coordinate $\s$ (representing the circle at infinity) is compact. The null direction $\t$, also corresponding to the time direction in the dual field theory, is non-compact. There exists another representation known as the plane representation, where both the directions on the plane $(x,\t)$ are non-compact.
 \be{}
 L_n=-x^{n+1}\p_x-(n+1)x^n \t \p_\t, ~~ M_n=x^{n+1} \p_\t.
 \ee
 The plane-to-cylinder mapping takes the following coordinate transformations,
 \be{}
 \t=i t e^{i \s},~~ x=e^{i \s}.
 \ee
In general, a finite BMS transformation takes the form \cite{Barnich:2012xq}
\be{}
\tilde{\s}=f(\s),~~ \tilde{t}=f^\prime(\s)t+\alpha(\s).
\ee

\subsection{Short review of the representation theory}
Now, we will take a brief look into the representation theory of $2d$ BMS$_3$ invariant field theories (BMSFT). First, let us summarise the highest weight representation in $2d$ CFTs. In $2d$ CFTs, the eigenvalues $h,\bar{h}$ of the operators $\mathcal{L}_0,\bar{\mathcal{L}}_0$ label the states,
\be{}
\mathcal{L}_0|h,\bar{h}\rangle=h|h,\bar{h}\rangle, ~~\bar{\mathcal{L}}_0|h,\bar{h}\rangle=\bar{h}|h,\bar{h}\rangle.
\ee
The CFT primaries are defined as
\be{}
\mathcal{L}_n|h,\bar{h}\rangle_p=\bar{\mathcal{L}}_n|h,\bar{h}\rangle_p=0~~\forall n>0.
\ee
The spectrum is bounded from below, and the CFT highest weight module is constructed by acting raising operators $\mathcal{L}_{-n},\bar{\mathcal{L}}_{-n}~(\forall n>0)$ on the primaries. The primary states are the basis of the highest weight representation, and the descendant states are obtained by acting with arbitrary numbers of raising operators on them. 

\medskip
\noindent
Now, we construct the highest weight representation of BMS in a similar fashion. We label the BMS states by the eigenvalues of dilatation and boost, $L_0$ and $M_0$, respectively.
\be{bmspri}
L_0 |\Delta,\xi\rangle=\Delta |\Delta,\xi\rangle,~~ M_0|\Delta,\xi\rangle=\xi |\Delta,\xi\rangle.
\ee
The BMS primary is defined by assuming the spectrum is bounded from below.
\be{}
L_n |\Delta,\xi\rangle_p= M_n|\Delta,\xi\rangle_p=0,~~ \forall n>0.
\ee
Similar to $2d$ CFT, the BMS highest weight modules are constructed by acting the raising operators $L_{-n}, M_{-n}~(\forall n>0) $ on the primaries.

\medskip
\noindent
The Virasoro states $| h,\bar{h}\rangle$ goes over to the BMS states  $| \D,\xi\rangle$ under the UR (Carrollian) contraction  \eqref{urbmsgen} \cite{Bagchi:2020rwb}
\bea{}
\non (\mathcal{L}_0-\bar{\mathcal{L}}_0)| h,\bar{h}\rangle &\implies& L_0| \D,\xi\rangle =\D |\D,\xi\rangle, \text{with~}\D=h-\bar{h},\\
(\mathcal{L}_0+\bar{\mathcal{L}}_0)| h,\bar{h}\rangle &\implies& M_0| \D,\xi\rangle =\xi |\D,\xi\rangle, \text{with~}\xi=\e(h+\bar{h})
\eea
with the following identifications $\D=h-\bar{h},\:\:\xi=\e(h+\bar{h})$. However, it is interesting to point out here that the UR contraction \eqref{urbmsgen} of the Virasoro highest weight representation does not end up giving the BMS highest weight module. It is easier to see this from the conditions on the primaries. 
\bea{}
\non \mathcal{L}_{n}|h,\bar{h}\rangle&=&(L_n+\frac{1}{\e}M_n)|h,\bar{h}\rangle=0,\:\:\forall n>0,~~\implies M_n|\D,\xi\rangle=0, \forall n>0,\\
\bar{\mathcal{L}}_{n}|h,\bar{h}\rangle&=&(-L_{-n}+\frac{1}{\e}M_{-n})|h,\bar{h}\rangle=0,\:\:\forall n>0,~~\implies M_
{-n}|\D,\xi\rangle=0, \forall n>0.
\eea
Thus, the UR contractions on the Virasoro highest weight representation give the following conditions
\be{}
L_0|\D,\xi\rangle=\D|\D,\xi\rangle,~~M_0 |\D,\xi\rangle=\xi|\D,\xi\rangle,~~M_n |\D,\xi\rangle=0 ~\forall n\neq0\,.
\ee
This is known as the induced representation of BMS in literature, explored in detail in \cite{Barnich:2014kra,Barnich:2015uva,Campoleoni:2016vsh}. Even though the BMS's highest weight and induced representations have very different characteristics, the operations of $L_0, M_0$ on the primaries are the same for both.\footnote{In our calculation, we have only used the action of $L_0, M_0$ on the BMS primaries $|\D,\xi\rangle$. Hence, our results are valid in both choices of representations.}

\bigskip
\noindent
However, it is possible to reach the BMS highest weight representation from Virasoro's highest weight through a different limit, reproducing the same algebra in \eqref{bms3al}. This is known as the non-relativistic (NR) contraction. This is opposite to the UR contraction described above, in the sense that the light cone opens up and the speed of light $c\to \infty$. The NR contraction on the Virasoro generators is realised as,
\be{nrlim}
L_n=\mathcal{L}_n+\bar{\mathcal{L}}_n, \:\: M_n=\e(\mathcal{L}_n-\bar{\mathcal{L}}_n)\,.
\ee
The readers can find more details on the NR contraction and the differences between the two singular limits in \cite{Bagchi:2009my,Bagchi:2019xfx}. In the non-relativistic limit, the CFT primary goes over to the BMS primaries $|\Delta,\xi \rangle$, with the identifications $\Delta=h+\bar{h},\:  \xi=\e(h-\bar{h})$.
\bes{}
\begin{align}
(\mathcal{L}_0+\bar{\mathcal{L}_0})|h,\bar{h}\rangle&=(h+\bar{h})|h,\bar{h}\rangle, \Rightarrow L_0|\Delta,\xi\rangle=\Delta|\Delta,\xi\rangle, \text{~with~} \Delta=h+\bar{h},\\
(\mathcal{L}_0-\bar{\mathcal{L}_0})|h,\bar{h}\rangle&=(h-\bar{h})|h,\bar{h}\rangle, \Rightarrow M_0|\Delta,\xi\rangle=\xi|\Delta,\xi\rangle, \text{~with~} \xi=\e(h-\bar{h})\,.
\end{align}\ees
The NR limit on the CFT algebra produces the Galilean Conformal Algebra (GCA). In terms of spacetime coordinate, the NR limit reflects as
\be{}
t\to t,~~x^i\to \e x^i, \implies, \frac{v}{c}\sim \frac{1}{c}\frac{x}{t}\to 0,~~\text{equivalently,~} c\to \infty\,.
\ee

\medskip
\noindent
It is also fascinating to note that the isomorphism between the CCA and GCA exists only when the boundary spacetime dimensions ($d$) is two. In $2d$, we contract only one direction; the temporal one for the UR (Carrollian) contraction and the spatial direction for the NR (Galilean) one (as opposed to more than one contracted spatial direction for $d>2$), reproducing the same algebra \eqref{bms3al}\cite{Bagchi:2019xfx}. We will exploit this isomorphism in a later part of the paper. 

\section{Revisiting circuit complexity construction  for Virasoro groups}\label{vircom}
After briefly reviewing the necessary details of the BMS$_3$ algebra and its contractions, we now focus on the other component of this paper, the circuit complexity. In this section, we revisit circuit complexity for the relativistic cousin of the BMS group, the infinite-dimensional Virasoro group. This review will come in handy before we dive into the details of BMS complexity. Note that, here, we are only considering one copy of the Virasoro algebra (the holomorphic part) similar to \cite{Caputa:2018kdj,Erdmenger:2020sup}. The contribution from the anti-holomorphic part comes in a similar fashion, and we will address the same in the next section. First, let us discuss the modifications necessary to use the notion of Nielsen's geometric approach \cite{nielsen2005geometric,nielsen2006quantum,dowling2006geometry} in computing the circuit complexity for an continuum field theory \cite{Caputa:2018kdj,Erdmenger:2020sup}.  

\medskip
\noindent
The geometric approach of Nielsen's complexity counts the discrete gates to reach from a reference to a target state, resulting in the construction of a path in the space of the unitary operators and then extremizes it \cite{nielsen2005geometric,nielsen2006quantum}. In general, one also needs to associate a penalty factor with various gates, and then perform the extremization of complexity functional. This extremization essentially gives geodesics on the symmetry group manifold. The penalty factor determines how hard it is to perform the given set of transformations. In our case, the gates are built out of the symmetry transformation generators for a continuous symmetry group. Hence, all gates are equally difficult to perform. Thus, the notion of the penalty factor does not bear any contribution (other than an overall pre-factor) in the complexity functional. Due to this choice, we can reach the target states that are only related to the reference state by group transformations. Also, the complexity depends on the choice of the reference state. 

\medskip
\noindent
Now, let us briefly review the circuit complexity for the Virasoro group following \cite{Caputa:2018kdj,Erdmenger:2020sup}. The Virasoro group is denoted by the group elements $(f(\s),\mathfrak{a})$. Here, $f(\s)$ with $\s \in \mathbb{S}^1$ denotes  diffeomorphisms on the circle $\mathbb{S}^1$. In other words, considering $\s$ as the light cone coordinate, $\s \to f(\s)$ gives the conformal transformations in $2d$. The extra contribution  $\mathfrak{a}\in \mathbb{R}$ comes from the central extension $c$ of the Virasoro group. We will first focus on the contribution from the conformal transformations $f(t,\s)$. 

\medskip
\noindent
Let us start with a reference state $|\psi_R\rangle$ of our choice, and the target state $|\psi_T\rangle$ is related to the reference state by unitary transformations of the group. 
\be{}
|\psi_{T}\rangle=U_{f(T)}|\psi_R\rangle\,.
\ee
Here, $U_{f(t)}$ denotes the unitary group transformations after time $t$. At $t=0,$ the initial transformation $U_{f(t=0)}=\mathrm{1}$ denotes the reference state itself, and $U_{f(t=T)}$ gives the final target state. Thus, we can write the finite transformation $U_{f(T)}$ in terms of infinitesimal group transformations
\be{}
U_{f(T)}=U_{\e(T)}U_{\e(T-dT)}\cdots U_{\e(dt)}\mathrm{1}.
\ee
Therefore, the infinitesimal symmetry generators play the role of unitary gates of Nielsen's complexity. In $2d$, the conserved energy-momentum tensor $T(\s)$ of the Virasoro group can be used to write the gates,
\be{}
Q(t)=\frac{1}{2\pi}\int d\sigma \: \epsilon(t,\sigma)\: T(\sigma).
\ee
Here, the gates $Q(t)$ are path ordered in the sense that earlier gates are applied first, and $\epsilon(t,\s)$ is the instantaneous group velocities. The geometric notion of Nielsen's complexity defines a path (geodesic, after extremization) in the manifold of the group transformations. Two infinitesimally close points on the path are related to each other by \cite{Caputa:2018kdj},
\be{}
U(t+dt)=e^{-Q(t)dt}\:U(t), ~~\text{where,\:\:} U_{f(t)}=U(t).
\ee
Now, we calculate the instantaneous group velocities $\e(t,\s)$ from the group elements (diffeomorphisms) $f(t,\s)$.
\be{}
f(t+dt,\sigma)=e^{\e(t,\sigma)dt}\cdot f(t,\sigma)
\ee
Here, `$\cdot$' refers to the group operation of Virasoro group $(f_1\cdot f_2=f_1 \circ f_2)$. Expanding upto the first order on both sides,
\bea{}
\non f(t,\s)+dt\:\frac{\p f(t,\s)}{\p t}&=&(\mathrm{1}+dt\:\e(t,s) ) \circ f(t,\s).\\
\non \implies \e(t,f(t,\s))&=&\frac{\p f(t,\s)}{\p t}.
\eea
Our aim is to extract the group velocity $\e(t,\s)$. To do so, we define the inverse diffeomorphism $F(t,\s)$ at each point to the diffeomorphisms $f(t,\s)$, such that
\be{}
F(t,f(t,\sigma))=\s.
\ee
Thus, 
\be{}
\e(t,\s)=\frac{\p f(t,F(t,\s))}{\p t}.
\ee
To find the expression of $\e(t,\s)$ we use the following
\be{elcal}
\frac{\p\: F(t,f)}{\p t}+\frac{\p F(t,f)}{\p f(t,\s)}\frac{\p\:f(t,\s)}{\p t}=0.
\ee
Next, we change the variable from $f(t,\s)$ to $\s$. Thus,
\bea{}
\non &&\frac{\p\: F(t,\s)}{\p t}+\frac{\p F(t,\s)}{\p \s}\frac{\p\:f(t,F(t,\s))}{\p t}=0.\\
&&\implies \frac{\p f(t,F(t,\s))}{\p t}=-\frac{\frac{\p\:F(t,\s)}{\p t}}{\frac{\p\:F(t,\s)}{\p\s}}.
\eea
This evaluates the group velocity in terms of the inverse diffeomorphism $F(t,\s)$ \cite{Caputa:2018kdj,Erdmenger:2020sup}
\be{}
\e(t,\s)=-\frac{\dot{F}(t,\s)}{F^\prime(t,\s)}.
\ee
Here, `$~\dot{}~$' and `$~\prime~$' denotes the derivative wrt $t$ and $\sigma$.

\medskip
\noindent
After evaluating the group velocity, we are left with writing the cost functional $\mathcal{F}$. We adapt Nielsen's approach to write the cost functional of a finite-dimensional system and modify it by imposing state-dependence in order to account for an infinite dimensional system. For this purpose, we use the density matrix $\rho(t)$, which is evolved from the initial density matrix $\rho_0$ by the symmetry transformations $U(t)$.
\be{}
\rho(t)=U(t)\rho_0U^{\dagger}(t),~~ \rho_0=|\psi_R\rangle\langle\psi_R|
\ee
We chose the one-norm cost function with the explicit state-dependence according to \cite{Caputa:2018kdj,Erdmenger:2020sup},
\be{costfunc1}
\mathcal{F}=|tr[\rho(t)Q(t)]|=|\langle \psi_R| U^{\dagger}(t)Q(t)U(t)|\psi_R\rangle|=\int d\sigma\: \epsilon(t,\sigma)\langle \psi_R| U^{\dagger}(t)T(\sigma)U(t)|\psi_R\rangle.
\ee
 There are a plethora of choices for cost-functional. For more details of possible choices for cost-functional interested readers are referred to \cite{nielsen2005geometric,Guo:2018kzl}. It is worth reminding here that the issue of penalty factors associated with the possible paths is already taken care of by choosing symmetry transformations as the gates $Q(t)$. Hence, the complexity functional $\mathcal{C}[f]$ gives the total cost of a constructed path from the reference to the target state in the group manifold in terms of group element $f(t)$.
\be{comfunc1}
\mathcal{C}[f]=\int dt \mathcal{F}=\frac{1}{2\pi}\int dt \int d\sigma\: \epsilon(t,\sigma)\langle \psi_R| U^{\dagger}(t)T(\sigma)U(t)|\psi_R\rangle\,.
\ee
Minimizing the complexity functional $C[f]$ gives the equation of motion in terms of the group path $f(t)$, the solution of which gives the optimized circuit. We then put the solution back into the complexity functional to obtain the on-shell value of the complexity arising from the optimal circuit starting from the chosen reference to the desired target state. For completeness, we show the construction of the complexity functional explicitly for the Virasoro group \cite{Caputa:2018kdj,Erdmenger:2020sup}.

\medskip
\noindent
First, we write the transformation law of the energy-momentum tensor $T$ under the conformal transformation in terms of the inverse diffeomorphism $F$, consisting of the usual Schwarzian $\{F,\s\}$.
\be{}
U_f^\dagger T U_f=F^{\prime 2}T(F)+\frac{c}{12}\{F,\s\},~~\text{where,\:}\{F,\s\}=\frac{F^{\prime\prime\prime}(\s)}{F^\prime(\s)}-\frac{3}{2}\left(\frac{F^{\prime\prime}(\s)}{F^\prime(\s)}\right)^2\,.
\ee
We also choose the reference state to be the CFT primary, $|\psi_R\rangle=|h\rangle$. Putting the transformation of $T$ and the expression for the group velocity $\e(t,\s)$ back into \eqref{comfunc1}, we write the complexity functional $\mathcal{C}[F]$ in terms of the inverse diffeomorphism, 
\be{compfunc1}
\mathcal{C}[F]=\int dt \:\mathcal{F}=\frac{1}{2\pi} \int dt \int d\s \left[-\dot{F}F^\prime \langle h|T(F)|h\rangle-\frac{c}{12}\frac{\dot{F}}{F^\prime}\{F,\s\} \right]\,.
\ee
In terms of the diffeomorphism $f(t,\s)$, the complexity functional takes the form
\be{}
\mathcal{C}[f]=\frac{1}{2\pi}\int^T_0 dt \int_0^{2\pi} d\s \frac{\dot{f}}{f^\prime} \left(-\langle h|T(\s)|h\rangle +\frac{c}{12}\{f,\s\} \right)\,.
\ee
Next, we will be accounting for the contribution due to the central extension $\mathfrak{a}$ in the complexity.

\paragraph{Contribution due to the central extension:} In \cite{Caputa:2018kdj,Erdmenger:2020sup}, the instantaneous velocity $\epsilon(t,\s)$ is identified with the non-centrally extended Maurer-Cartan form $\theta_{f^{-1}}$ of the Virasoro group. This can be easily understood by writing the symmetry transformation $f(t)$ of the Virasoro group in terms of group velocity $\e(t,\s)$ and its infinitesimal close point on the path of unitary manifold, followed by multiplying both sides with the inverse element $F$, and then taking derivative,
\be{mc1}
 f(t)=e^{\int_t^s \:\e{(s^\prime)ds^\prime}}f(s) ,\quad
\implies  \frac{d}{ds}\Big|_{s=t} (f(t)\cdot F(s))=-\e(t).
\ee
The non-centrally extended Maurer-Cartan form of the Virasoro group in terms of $f^{-1}$ (rather than $f$) is the group velocity,
\be{}
\theta_{f^{-1}}=\frac{d}{ds}\Big|_{s=t} (f(t)\circ F(s))=-\e(t).
\ee

\medskip
\noindent
However, in \cite{Erdmenger:2020sup}, the analysis was carried forward by arguing that it is also necessary to use the generalized (centrally extended)  Maurer-Cartan form in order to find the complete contribution of the central extension $c$ in the circuit complexity. The authors argued not only to use the path through the group transformations $f(t)$, but also the path through the real number $\mathfrak{a}(t)$. This statement translates to the generalized version of \eqref{mc1}, 
\be{}
(f(t),\mathfrak{a}(t))=e^{\int_t^s \:(\e{ (s^\prime),\:\beta(s^\prime) )ds^\prime}} \big(f(s),\mathfrak{a}(s) \big)\,. 
\ee
Multiplying from the right side by $ \big(f,\mathfrak{a} \big)^{-1}=(F,-\mathfrak{a})$ and taking the derivative wrt $s$, we obtain the Maurer-Cartan form $\theta$ and its central extension $m_\theta$ with respect to the inverse element $f^{-1}$ of the Virasoro group,
\bea{thmth}
\big(\theta,m_{\theta} \big)_{f^{-1}}=-\big(\e(t),\beta(t) \big)&=& \frac{d}{ds} \Big( \big(f(t),\mathfrak{a}(t))\cdot(F(s),-\mathfrak{a}(s))\Big),\\
\non &=&\Bigg( \frac{d}{ds}\Big|_{s=t} (f(t)\circ F(s)), \:\: \frac{d}{ds}\Big|_{s=t}\mathfrak{C}\big(f(t),F(s)\big)+a(t)-a(s) \Bigg)\,.\\
\eea
Here, in the last line, we have used the group multiplication rule of the Virasoro group
$(g_1,\a_1)\cdot(g_2, \a_2)=(g_1\circ g_2,\a_1+\a_2+\mathfrak{C}(g_1,g_2))$. Also, the real number valued part is assumed to be constant $\mathfrak{a}(t)=\mathfrak{a}(s)=$const. The term $\mathfrak{C}\big(f(t),F(s)\big)$ is the 2-cocycle defining the central extensions. The centrally extended Maurer-Cartan form takes the form for the Virasoro group \cite{Erdmenger:2020sup,Oblak:2017ect},
\be{thmth2}
(\theta,m_{\theta})=\Bigg(\frac{\dot{F}}{F^\prime},\frac{1}{48\pi} \int_0^{2\pi}d\sigma \frac{\dot{F}}{F^\prime}\:\Big(\frac{F^{\prime\prime}}{F^\prime}\Big)^{ ^\prime}\:\Bigg)\,.
\ee
This begs the following modification in the cost function \eqref{costfunc1} due to the velocities $\e(t)$ and $\beta(t)$ of centrally extended Virasoro group,
\be{costfunc2}
\mathcal{F}=c \beta(t) \:\: + \:\: \int d\sigma\: \epsilon(t,\sigma)\langle \psi_R| U^{\dagger}(t)T(\sigma)U(t)|\psi_R\rangle .
\ee
Putting both the velocities $\e(t)=-\theta$ and $\beta(t)=-m_\theta$ from \eqref{thmth2}  back in the complexity functional \eqref{compfunc1}, we find
\bea{compfunc2}
\non \mathcal{C}[F]&=&\int dt \:\mathcal{F}\\
\non &=& \int dt \int d\s \left[-\dot{F}F^\prime \frac{\langle h|T(F)|h\rangle}{2\pi}+\frac{c}{24\pi}\frac{\dot{F}}{F^\prime}\{F,\s\} -  \frac{c}{48\pi} \frac{\dot{F}}{F^\prime}\:\Big(\frac{F^{\prime\prime}}{F^\prime}\Big)^{ ^\prime}\right]\\
&=& \int d\s\: dt \left[ -j_0(F)\dot{F}{F^\prime}+\frac{c}{48\pi} \frac{\dot{F}}{F^\prime} \:\Bigg(\frac{F^{\prime\prime\prime}}{F^\prime}-2 \Big(\frac{F^{\prime\prime}}{F^\prime} \Big)^{ ^2}\Bigg)\right].
\eea
Here, $j_0$ is a constant. The complexity functional \eqref{compfunc2}, including the contribution from the central term, is equal to the geometric action for the Virasoro group \cite{Alekseev:1988ce,Alekseev:2018pbv,Alekseev:1990mp,Barnich:2017jgw}. This is one of the central claims of \cite{Erdmenger:2020sup}, where the authors show the equivalence of the complexity functional with the on-shell value of the geometric action for centrally extended direct product groups, e.g. Virasoro and Kac-Moody. To construct the circuit complexity for symmetry groups in the case of non-Lorentzian field theories, we start our analysis from here in the next section.\footnote{For extremization of the complexity functional to find the optimal path $f$ and the corresponding value of the complexity, readers are referred to \cite{Erdmenger:2020sup}.}

\section{From CFT to BMS complexity}\label{liman}
In the last section, we briefly familiarised ourselves with the formulation of circuit complexity using symmetry transformations as the unitary gates for the infinite-dimensional centrally extended Virasoro group. Strikingly, the modified cost function (due to the contribution from the central extension \eqref{costfunc2}) gives the complexity functional \eqref{compfunc2}, which is equivalent to the geometric action of the centrally extended Virasoro group. Now, we will translate the above formulation for the BMS$_3$ group ($2d$ Carrollian Conformal group) and find the circuit complexity. We will approach the problem in two ways: first, as a limit to the formulation used in the context of the Virasoro group in section \ref{vircom}; and then using the intrinsic symmetry transformations of BMS.

\medskip
\noindent
BMS$_3$ is the asymptotic symmetry group at the boundary of $3d$ asymptotically flat spacetime. The null boundary has a structure $\mathbb{R}\times \mathbb{S}^1$. BMS$_3$ group is the semi-direct product of the local co-ordinate transformations (diffeomorphisms) on the circle at $\mathscr{I}^+$ (also, known as the superrotations) and the angle-dependent supertranslations. Also, $c_L$ and $c_M$ are the central charges in the $[L,L]$ and $[L,M]$ brackets in \eqref{bms3al}.
\be{}
 \text{BMS}_3=\underbrace{\text{Diff}(S^1)}_{\text{Super-rotation}} \ltimes \underbrace{\text{Vec}(S^1)}_{\text{Supertranslation}}\,.
 \ee
 The elements of the BMS$_3$ groups are the diffeomorphisms $f(t,\s)$ with central extension $c_1$ and the supertranslations $\a(t,\s)$ with central extension $c_2$. They are given by the periodicity conditions: 
 \bea{}
 f(\s+2\pi)&=&f(\s)+2\pi,~~ f^\prime(\s)>0,\\
 \a(\s+2\pi)&=&\a(\s).
 \eea

\smallskip
\noindent
In section \ref{bmsrevisit}, we showed that it is possible to reach the BMS$_3$ algebra starting from the Virasoro algebra. Hence, we will first use the limiting analysis to find the BMS$_3$ circuit complexity from the analogous results of the centrally extended Virasoro group. We start with the Virasoro cost functional \eqref{costfunc1} without the contribution from the central extension,
\be{vircompc1}
\mathcal{C}_1=\int dt \:\mathcal{F}=\frac{1}{2\pi}\int d t \int d\sigma\: \epsilon(t ,\sigma)\big\langle \psi_R\big| U^{\dagger}(t )\left
(\:T+\bar{T}\:\right)U(t )\big|\psi_R\big\rangle\,.
\ee
\be{}
\mathcal{C}=\mathcal{C}_1+\mathcal{C}_2\,,
\ee
where $\mathcal{C}_2$ is the contribution from the central extension to the circuit complexity $\mathcal{C}$, which we discuss later. Here, we have also explicitly written the contribution from the holomorphic and anti-holomorphic parts $T,\bar{T}$ of the stress tensors for $2d$ CFT. Now, we use the non-relativistic limit \eqref{nrlim} on the CFT stress tensors.\footnote{In $2d$, the non-relativistic and the ultra-relativistic contractions of the Virasoro algebra gives the GCA and the CCA, respectively, which are isomorphic to the BMS$_3$ algebra. The isomorphism between GCA and CCA is not seen in $d>2$. } For the non-relativistic limit the contraction is given by \cite{Bagchi:2015nca},
\be{}
T+\bar{T}=T_1,~~ \lim_{\e\to0}\e(T-\bar{T})=T_2.
\ee
Here, $T_1, T_2$ are the stress tensors for BMS$_3$. The reference state for $2d$ CFT is chosen to be the primary $|h, \bar{h}\rangle$. In the non-relativistic limit, the CFT primary goes to the BMS primaries $|\Delta,\xi \rangle$ \cite{Bagchi:2020rwb}. Thus, 
\be{}
|\psi_R\rangle=|h, \bar{h}\rangle \xRightarrow[\text{limit}]{\text{NR}} |\Delta,\xi \rangle\,.
\ee
Hence the contribution from \eqref{vircompc1} in the non-relativistic limit gives,
\be{bmscomp2}
\mathcal{C}_1=\int dt \:\mathcal{F}=\frac{1}{2\pi}\int d t \int d\sigma\: \epsilon(t ,\sigma)\big\langle \D,\xi\big| U^{\dagger}(t )T_1(t ,\s)U(t )\big|\D,\xi\big\rangle\,.
\ee
Next, we expand the BMS$_3$ stress tensors with modes on the cylinder \cite{Bagchi:2015nca},
\be{}
T_1(t ,\s)=\sum_n \left(L_n-i n\s M_n \right)e^{-i n t }+\frac{c_L}{12}\delta_{n,0},~~~
T_2(\t,\s)=\sum_n M_n e^{-i n t }+\frac{c_M}{12}\delta_{n,0}.
\ee
Here, $L_n$ and $M_n$'s are generators of the BMS$_3$ algebra in \eqref{bms3al}. Thus,
\be{comfunc2}
\mathcal{C}_1=\frac{1}{2\pi}\int d t \int d\sigma\: \epsilon(t ,\sigma) \sum_{n}\big\langle \D,\xi \big| U^{\dagger}(t )\Big(\left(L_n-i n\s M_n \right)e^{-i n t }+\frac{c_L}{12}\delta_{n,0}\Big)U(t )\big|\D,\xi\big\rangle
\ee
On the cylinder, the generators are written in terms of the conserved currents $j^{cyl}$ and $p^{cyl }$  \cite{Jiang:2017ecm},
\be{}
L_n^{cyl}=-\frac{1}{2\pi}\int_0^{2\pi} d\s e^{in\s}j^{cyl}(\s),~~M_n=-\frac{1}{2\pi}\int_0^{2\pi} d\s e^{in\s}p^{cyl }(\s)
\ee
Under the finite BMS transformations,
\be{}
\tilde{\s}=F(\s),~~\tilde{t }=t  F^\prime(\s)+\a(\s)
\ee
the transformation relations of the conserved currents are \cite{Basu:2015evh,Jiang:2017ecm} 
\bes{}
\label{concur1}
\begin{align}
\tilde{p}(\s)&=F^\prime(\s)^2 p(\tilde{\s})+\frac{c_M}{12}\{F,\s\},\\
\tilde{j}(\s)&=F^{\prime}(\s)^2 j({\tilde{\s}}) +2 F^\prime(\s) \a^\prime(\s) p(\tilde{\s})+F^\prime(\s)^2 \a \: \partial_{\tilde{\s}}p(\tilde{\s})+\frac{c_L}{12}\{F,\s \}+\frac{c_M}{12} \llbracket (F,\a ),\s \rrbracket\,.
\end{align}
\ees
Here, prime `$\: \prime$\:' denotes the derivative wrt $\s$. Also, we have the Schwarzian $\{F,\s\}$ and the BMS Schwarzian $ \llbracket (F,\a ),\s \rrbracket$, 
\bes{}
 \begin{align}
 \{F,\s\}&=\frac{F^{\prime\prime\prime}}{F^{\prime}}-\frac{3}{2}\left(\frac{F^{\prime\prime}}{F^\prime} \right)^2,\\
  \llbracket (F,\a ),\s \rrbracket&=\frac{[3(F^{\prime\prime})^2-F^\prime F^{\prime\prime\prime}]\a ^\prime-3F^\prime F^{\prime\prime}\a ^{\prime\prime}+(F^\prime)^2\a ^{\prime\prime\prime}}{(F^\prime)^3}\,.
 \end{align}
 \ees
 We choose the reference state to be a BMS primary $|\Delta,\xi\rangle$ \eqref{bmspri}. Hence, the only mode that will contribute is $n=0$, or $L_0$ in the sandwich between two BMS primaries \eqref{comfunc1}. We are rewriting the transformation relation of the currents \eqref{concur1}  in a slightly different notation following \cite{Merbis:2019wgk,Barnich:2017jgw}.
 \bes{}
 \label{concur2}
 \begin{align}
 \tilde{p}(\s)&=F^\prime(\s)^2p(F)-\frac{c_2}{24\pi}\{F,\s\},\\
 \tilde{j}(\s)&=F^\prime(\s)^2 \Big( \partial_F p(F) \a(F)+2 \partial_{F} \a(F)p(F)-\frac{c_2}{24\pi}\partial_F^3 \a(F) \Big)+F^\prime(\s)^2j(F)-\frac{c_1}{24\pi}\{F,\s \}\,.
\end{align}
 \ees
 Here, $\prime$ denotes the derivatives wrt $\s$. Also, the BMS central charges are $c_1=-2\pi c_L , c_2=-2\pi c_M $. The BMS Schwarzian is reflected in the term $\partial_F^3 \a(F)$,
\be{}
\partial_F^3 \a(F)=\frac{ \llbracket (F,\a ),\s \rrbracket}{(F^\prime)^5}.
\ee 
In the limiting analysis, we find that only the transformation of the current $j(\s)$ (corresponding to the superrotations) contributes to the unitary transformation $U^\dagger T_1(\s,t ) U$. In terms of the inverse diffeomorphisms, 
\be{}
F(t,f(t,\s))=\s
\ee
the group velocity $\epsilon(t,\s)$ is
\be{}
\e(t,\s)=-\frac{\dot{F}(t,\s)}{F^\prime(t,\s)}
\ee
Finally, using \eqref{concur2} we can rewrite the complexity functional \eqref{comfunc1} as,
\bea{}
\label{comfunc3}
\non \mathcal{C}_1&&=\frac{1}{2\pi}\int d t \int d\sigma\: \Big[ -\dot{F}F^\prime \left(j_0(F)+ \partial_F p_0(F) \a(F)+2 \partial_{F} \a(F)p_0(F)-\frac{c_2}{24\pi}\partial_F^3 \a(F)\right)\\
&& ~~+\frac{c_1}{24\pi}\frac{\dot{F}}{F^\prime}\left( \frac{F^{\prime\prime\prime}}{F^{\prime}}-\frac{3}{2}\left(\frac{F^{\prime\prime}}{F^\prime} \right)^2\right)\Big]\,.
\eea
Here, $j_0$ and $p_0$ are two constants.
\paragraph{Contribution from the central extensions:} Similar to the arguments discussed in section \ref{vircom}, we also add here the contribution due to  the central extension $c_1$,\footnote{We do not add here the other central term $c_2$ due to absence of the supertranslation current $p(\s)$ in \eqref{bmscomp2}. }
 \be{}
 \mathcal{C}_2=\frac{1}{2\pi}\int d t \int d\sigma\: \left(-\frac{\dot{F}}{F^\prime}\right)\:\left[\left(\frac{c_1}{48\pi} \right)\left(\frac{F^{\prime\prime}}{F^\prime} \right)^\prime  \right]\,.
 \ee
 Thus, the complexity functional of BMS$_3$ as a limit of the Virasoro is given as,
 \bea{compfunc3}
\non \mathcal{C}&=&\mathcal{C}_1+\mathcal{C}_2\\
\non  &=& \frac{1}{2\pi}\int d t \int d\sigma\: \Big[ -\dot{F}F^\prime \left(j_0(F)+ \partial_F p_0(F) \a(F)+2 \partial_{F} \a(F)p_0(F)-\frac{c_2}{24\pi}\partial_F^3 \a(F)\right)\\ && ~~+\frac{c_1}{24\pi}\frac{\dot{F}}{F^\prime}\left( \frac{F^{\prime\prime\prime}}{F^{\prime}}-\frac{3}{2}\left(\frac{F^{\prime\prime}}{F^\prime} \right)^2- \frac{1}{2}\left(\frac{F^{\prime\prime}}{F^\prime} \right)^\prime\right)\Big]
\eea
 The terms in the last line of \eqref{compfunc3} with the central charge $c_1$ can be expressed as,
 \bea{compfunc4}
 \non
\int d t d\sigma\: \frac{c_1}{48\pi}\frac{\dot{F}}{F^\prime}\left(\frac{F^{\prime\prime\prime} }{F^\prime}-2\left(\frac{F^{\prime\prime}}{F^\prime} \right)^2  \right)
 &=& \int d t  d\sigma\: \frac{c_1}{48\pi}\frac{\dot{F}}{F^\prime}\Big( \left(\frac{F^{\prime\prime} }{F^\prime}\right)^\prime+F^{\prime\prime}\left(\frac{1}{F^\prime} \right)^\prime \Big)\\
 \non &=&- \int d t  d\sigma\: \frac{c_1}{48\pi} \Big(\frac{\dot{F}^{\prime\prime}}{F^\prime}\Big)+\frac{c_1}{48\pi}\int d t d\sigma\: \Bigg(\frac{\dot{F}}{F^\prime}\frac{F^{\prime\prime}}{F^\prime}+\frac{\dot{F}^\prime}{F^\prime} \Bigg)^\prime  \\
 \eea
 The second term in \eqref{compfunc4} is a total derivative term. So, the final form of the complexity functional is
 \bea{compfunc5}
\non  \mathcal{C}&=& -\frac{1}{2\pi}\int d t \:\: d\sigma\: \Big[ \dot{F}F^\prime j_0(F)+  \frac{c_1}{48\pi} \frac{\dot{F}^{\prime\prime}}{F^\prime}\\
 && ~+ \dot{F}F^\prime\Big( \partial_F p_0(F) \a(F)+2 \partial_{F} \a(F)p_0(F)-\frac{c_2}{24\pi}\partial_F^3 \a(F)\Big)
 \eea
 The final expression for BMS$_3$ circuit complexity \eqref{compfunc5} is equal to the geometric (co-adjoint orbit) action of  BMS$_3$ in literature \cite{Barnich:2017jgw,Merbis:2019wgk}.  This is the expression we obtain from the limiting analysis.
 
 \smallskip
 \noindent
Before moving into the next section, let us spare a moment here to understand \eqref{compfunc5}. The geometric action corresponds to generic (massive) co-adjoint orbits of BMS$_3$ when the orbit representatives $j_0$ and $p_0$ are constant and non-zero ($j_0 \neq 0, p_0 \geq -\frac{c_2}{48\pi}$). For, the Flat Space Cosmology (FSC) solutions \footnote{The Flat Space Cosmologies (FSC) are shifted boost orbifolds of $3d$ flat spacetimes \cite{Cornalba:2003kd}. Alternatively, they can be understood as the flat space limit (AdS radius $l\to \infty$) of the non-extremal BTZ black holes. As a consequence of this limit, the outer horizon of the BTZ black hole is pushed to infinity, whereas the inner horizon stays at a finite distance, known as the cosmological horizon. The nature of the spacelike and timelike coordinates are interchanged due to the limit \cite{Riegler:2014bia}. More details regarding the thermodynamics of these cosmologies can be found in \cite{Barnich:2012xq,Bagchi:2013lma,Detournay:2014fva}, and the BMS-Cardy formula in the context of flat space holography reproducing the Bekenstein-Hawking entropy for these can be found in \cite{Bagchi:2012xr}.)} we have $p_0 >0$ and $j_0\neq 0$. Whereas, for conical deficit solutions we have $-\frac{c_2}{48\pi}<p_0<0$ and $j_0 \neq 0$. Both of these two types (FSC and conical deficit solutions) of generic (massive) orbits of BMS$_3$ have their energy bounded from below \cite{Merbis:2019wgk}.

\medskip
\noindent
There exists another type of BMS$_3$ orbits, called the vacuum BMS$_3$ orbit (e.g Minkowski vaccuum with $j_0=0$ and $p_0=-\frac{c_2}{48\pi}$). Here also, the energy is bounded from below. But the geometric action on these vacuum BMS$_3$ co-adjoint orbits are different than \eqref{compfunc5} and we will not discuss them in our work. The readers are requested to go through \cite{Merbis:2019wgk} for more details on the vacuum BMS$_3$ orbits.

 \section{BMS$_3$ complexity: Intrinsic way}\label{intan}
In the previous section, we used the limiting analysis to find complexity functional for the BMS$_3$ group from the circuit complexity of the Virasoro group. In the limiting analysis, the coadjoint orbit action of BMS gives the complexity functional. However, the information of the transformation of $p(\s)$, the current corresponding to the supertranslation generators $M_n$, did not enter the complexity functional. It is always possible to add a Hamiltonian $\int\:dt\: H$, preserving the global symmetries of the theory to the geometric action on the coadjoint orbit. However, it does not seem convincing from the point of view of calculating the complexity functional only (if our sole intention is not to match the complexity functional with the geometric action). In this section, we try to modify our computation by taking the intrinsic approach. First, we calculate the complexity functional strictly using the symmetry generators of BMS$_3$. Later, we extremize the complexity functional and find the optimal path in the space of group transformations.

  \subsection{Constructing the BMS Complexity functional} We start with the general expression for cost functional using the conserved currents $J(t,\s)$ for symmetry gates $Q(t)=\frac{1}{2\pi}\int d\sigma \: \epsilon(t,\sigma)\: J(\sigma)$. Here, $\epsilon(t,\s)$ is the instantaneous velocity and the conserved currents for BMS supertranslations and superrotations are $j(\s),p(\s)$. Their transformation relations are given in \eqref{concur2}. Thus, the cost functional for BMS$_3$ takes the form,
\be{costfunc4} 
 \mathcal{F}=\int d\sigma\: \: \Big[\epsilon_L(t,\sigma)\langle \psi_R| U^{\dagger}(t)j(\sigma)U(t)|\psi_R\rangle+\epsilon_M(t,\sigma)\langle \psi_R| U^{\dagger}(t)p(\sigma)U(t)|\psi_R\rangle\Big]\,.
  \ee
  We have used two different instantaneous velocities $\e_L$ and $\e_M$ corresponding to the transformations $f(t,\s)$ (diffeomorphisms) and $\a(t,\s)$ (supertranslations). Thus, the complexity functional constructed from the cost functional \eqref{costfunc4} is,
 \bea{}
\non\mathcal{C}=\int dt \mathcal{F}&=&\frac{1}{2\pi}\int dt \int d\sigma\: \Big[\epsilon_L(t,\sigma)\langle \psi_R| U^{\dagger}(t)j(\sigma)U(t)|\psi_R\rangle+\epsilon_M(t,\sigma)\langle \psi_R| U^{\dagger}(t)p(\sigma)U(t)|\psi_R\rangle\Big],\\
&=& \frac{1}{2\pi}\int dt \int d\sigma\: \Big[ \epsilon_L(t,\sigma)\langle \psi_R|\tilde{j}(\s) |\psi_R\rangle+\epsilon_M(t,\sigma)\langle \psi_R| \tilde{p}(\s)|\psi_R\rangle\Big]\,.
\eea  
We are also using the inverse diffeomorphism $F(t,\s)$ at each point $f(t,\s)$ such that, 
 \be{}
 F(t,f(t,\s))=\s \,.
 \ee
Using the transformations of the currents \eqref{concur2}, 
 \bea{compfunc6}
\non \mathcal{C}[F]&=&\frac{1}{2\pi}\int dt \: d\sigma \Bigg[\e_M(t,\s) \Big( F^\prime(\s)^2p_0(F)-\frac{c_2}{24\pi}\{F,\s\}\Big)\\
 \non &&+\e_L(t,\s)\Big(F^\prime(\s)^2 \Big[ \partial_F p_0(F) \a(F)+2 \partial_{F} \a(F)p_0(F)-\frac{c_2}{24\pi}\partial_F^3 \a(F) \Big]+F^\prime(\s)^2j_0(F)-\frac{c_1}{24\pi}\{F,\s \} \Big) \Bigg].\\
 \eea
 Here, $p_0$ and $j_0$ are constants. Now, we will calculate the instantaneous velocities $\epsilon_L, \epsilon_M$. For this, we use the BMS group transformation rules. For a finite BMS transformation, we have
\be{}
\sigma \to f(\sigma),~~t\to tf^\prime(\sigma)+\alpha(\s)\,.
\ee
For two such consecutive transformations, we go from $(\s,t)$ to $(\s_2,t_2)$,
\be{}
(\s,t)\xrightarrow{(f_1,\a_1)}(\s_1,t_1)\xrightarrow{(f_2,\a_2)}(\s_2,t_2)
\ee
where,
\be{bmstrans1}
\s_1=f_1(\s),~~\text{and,~}\s_2=f_2(\s_1)=f_2(f_1(\s))=(f_2 \circ f_1)(\s).
\ee
Also,
 \bea{bmstrans2}
 \non t_1&=&t f^\prime_1(\s)+\a_1(\s),\\
 t_2&=&t_1 f^\prime_2(\s_1)+\a_2(\s_1)= t \frac{\partial (f_2\circ f_1)(\s)}{\p \s}+\a_1 \frac{\p (f_2\circ f_1)(\s)}{\p f_1(\s)}+\a_2 (f_1(\s))\,.
 \eea
Now, if we directly go from $(\s,t)$ to $(\s_2,t_2)$ the following also holds true.
 \bea{bmstrans3}
\non (\s,t)&\xrightarrow{(f_3,\a_3)}&(\s_2,t_2).\\
 \text{such that,~} \s_2&=f_3(\s)&,~~~~~ t_2=t \frac{\p f_3(\s)}{\p \s}+\a_3(\s).
 \eea
 Comparing, \eqref{bmstrans1} and \eqref{bmstrans2} with \eqref{bmstrans3}, we get the following transformation law of the group elements
\bea{groupop}
\non f_3(\s)&=&(f_2\circ f_1)(\s)=f_2(f_1(\s)).\\
\a_3(\s)&=&(\a_2 \circ f_1)(\s)+\a_1(\s)\frac{\p (f_2 \circ f_1)(\s)}{\p f_1(\s)}=\a_2 (f_1(\s))+\a_1(\s)\frac{\p f_2(f_1(\s))}{\p f_1(\s)}.
\eea
We will use \eqref{groupop} to find the instantaneous velocities. The symmetry transformation $(f(t,\s),\a(t,\s))$ for two infinitesimally close points in the path of the circuit is related as
 \be{}
\big(f(t+dt,\s),\a(t+dt,\s)\big)=e^{(\e_L(t,\s),\e_M(t,\s))dt}\:\:(f(t,\a),\a(t,\s)).
\ee
Expanding in the first order,
\bea{}
\non &\implies& \big(f(t,\a),\a(t,\s)\big)+dt\:\: (\p_t f (t,\s),\p_t \a(t,\s))=\big(\mathbb{I} + dt\:\:(\e_L(t,\s),\e_M(t,\s)) \big)\cdot \big(f(t,\s),\a(t,\s)\big)\\
&\implies& dt\:\: (\p_t f (t,\s),\p_t \a(t,\s))= dt \:\:\big(\e_L(t,\s)),\e_M(t,\s)\big)\cdot (f(t,\s),\a(t,\s))\,.
\eea
Here, `$\cdot$' represents the group operation in \eqref{groupop}.
Thus, we have
\bea{}
\p_t f(t,\s)&=&\e_L(t,f(t,\s)),\\ \label{em1}
\p_t \a(t,\s)&=&\e_M(t,f(t,\s))+\frac{\p \e_L(t,f(t,\s))}{\p f(t,\s)} \a(t,\s).
\eea
Now, we will use the inverse diffeomorphism $F(t,f(t,\s))=\s$ to find the instantaneous velocities $\e_L,\e_M$ at $(t,\s)$. Thus, 
 \be{el2}
 \e_L(t,\s)=\frac{\p f(t,F(t,\s))}{\p t}=-\frac{\dot{F}(t,\s)}{F^\prime(t,\s)}.
 \ee
 The calculation of $\e_L$ is similar to \eqref{elcal}. Similarly, we have from \eqref{em1},
\bea{em2}
\non \e_M(t,\s)&=&\p_t \a(t, F(t,\s))-\a(t,F(t,\s)) \frac{\p \e_L(t,\s)}{\p \s},\\
&=& \dot{\a}(t,F)+\a(t,F)\left(\frac{\dot{F}^\prime}{F^\prime}-\frac{\dot{F}F^{\prime \prime}}{F^{\prime^2}} \right).
\eea 
We use group velocities $\e_L(t,\s)$ \eqref{el2} and $\e_M(t,\s)$ \eqref{em2} in the complexity functional \eqref{compfunc5} to write 
 \bea{compin1}
\non \mathcal{C}[F]&=&\frac{1}{2\pi}\int dt \: d\sigma \Bigg[-\dot{F}F^\prime\Big[ \partial_F p_0(F) \a(F)+2 \partial_{F} \a(F)p_0(F)-\frac{c_2}{24\pi}\partial_F^3 \a(F)+j_0(F) \Big]+\frac{c_1}{24\pi}\frac{\dot{F}}{F^\prime}\{F,\s \} \\
&&~~+ \Bigg( \dot{\a}(t,F)+\a(t,F)\left(\frac{\dot{F}^\prime}{F^\prime}-\frac{\dot{F}F^{\prime \prime}}{F^{\prime^2}} \right)\Bigg)\Big( F^\prime(\s)^2p_0(F)-\frac{c_2}{24\pi}\{F,\s\}\Big)\Bigg].
 \eea
 Due to the presence of the central extension $c_1$ we also add the modification in the cost function similar to \eqref{costfunc2},
 \begin{align}
 \begin{split}
 & \mathcal{F}= \int d\sigma\: \Big[ \epsilon_L(t,\sigma)\langle \psi_R|\tilde{j}(\s) |\psi_R\rangle+\epsilon_M(t,\sigma)\langle \psi_R| \tilde{p}(\s)|\psi_R\rangle\Big]~~+c_1\beta(t)\,,\footnotemark\,\\&
  \mathcal{C}=\int dt \mathcal{F}, ~~~ \text{where,}~~\beta=-m_\theta=-\frac{1}{48\pi}\int_0^{2\pi} d\s \frac{\dot{F}}{F^\prime}\Big(\frac{F^{\prime\prime}}{F^\prime}\Big)^{ ^\prime} .
  \end{split}
 \end{align}
 \addtocounter{footnote}{0}
\footnotetext{The form of $\beta$ remains the same as the Virasoro case, in \eqref{thmth2}.}
\stepcounter{footnote}
 Thus, the final form for the complexity functional for BMS$_3$ is given by
  \bea{compin2}
\non \mathcal{C}[F]&=&\frac{1}{2\pi}\int dt \: d\sigma \Bigg[-\dot{F}F^\prime\Big[ \partial_F p_0(F) \a(F)+2 \partial_{F} \a(F)p_0(F)-\frac{c_2}{24\pi}\partial_F^3 \a(F)+j_0(F) \Big]-\frac{c_1}{48\pi}\frac{\dot{F}^{\prime\prime}}{F^\prime}\\
&&~~+ \Big( \dot{\a}(F)+\a(F)\left(\frac{\dot{F}^\prime}{F^\prime}-\frac{\dot{F}F^{\prime \prime}}{F^{\prime^2}} \right)\Big)\Big( F^\prime(\s)^2p_0(F)-\frac{c_2}{24\pi}\{F,\s\}\Big)\Bigg].
 \eea 
 The first line of \eqref{compin2} is the one obtained as a limit to Virasoro complexity functional in \eqref{compfunc5}. Putting, $\e_M=0$, we recover the geometric action on the generic (massive) coadjoint orbit for BMS$_3$ \cite{Barnich:2017jgw,Merbis:2019wgk}. However, the terms in the second line in \eqref{compin2} is not recovered by deforming the geometric action with the addition of Hamiltonian. We will comment on this in a moment. The second line is essential because it is not convincing to ignore the contribution from the supertranslation generators in the construction of the gates $Q(t)$. Next, we extremize the complexity functional \eqref{compin2} to obtain the equations of motions and solve for $f(t,\s),\a(t,\s)$. It is to note here, however, if we extremize the functional with $\e_M=0$, $\a$ remains undetermined. 
 
  \medskip
 \noindent
Before proceeding further, we connect the BMS complexity functional with some of the previous studies. Identifying 
$j_0,p_0,c_1$  in the following way and setting the instantaneous velocity $\e_M$ to be a constant,\footnote{There is no a priori reason to choose $\e_M(t,\s)$ as a constant for the complexity calculation.}
\be{}
p_0=k\frac{\mathfrak{M}_0}{2},\:\: j_0=k\mathfrak{L}_0, \:\:c_1=0, \:\: c_2=-2\pi c_M=24\pi k,\:\:\e_M=1.
\ee
Here we have used the fact, that for Einstein gravity $c_1$ can be set zero. Then we see that \eqref{compin2} becomes the deformed geometric action (by addition of Hamiltonian) of BMS$_3$ in \cite{Merbis:2019wgk}. Here,  $\mathfrak{M}_0$ and $\mathfrak{L}_0$ (the orbit representatives) are proportional to the angular momentum and mass of the gravitational saddle point. We use the notations from the authors \cite{Merbis:2019wgk} to write,
\bea{}
\non \mathcal{C}[F]&=&\mathcal{I}_{CS}\left[F,\a,\mathfrak{L}_0,\mathfrak{M}_0\right]\\
 &=&-\frac{k}{2\pi}\int dt\:d\s \: \: \Bigg[ \Big(\mathfrak{L}_0+\mathfrak{M}_0\partial_F \a(F)-\partial_F^3 \a(F)\Big) \dot{F}F^\prime-\frac{1}{2}\Big( \mathfrak{M}_0F^{\prime 2}-2\{F,\s\} \Big)\Bigg]. ~~~
 \eea 
This is the same as the effective two-dimensional action obtained from  bulk Chern-Simons classical gravity action for three-dimensional flat spacetime, after using the BMS$_3$ boundary conditions mentioned in \cite{Barnich:2006av}. Here, the boundary theory is equivalent to the coadjoint action on the geometric orbit of BMS$_3$ with non-trivial bulk holonomies.

 \subsection{Extremisation of the complexity functional} Now, we extremize the BMS complexity functional \eqref{compin2} wrt $f(t,\s),\a(t,\s)$. Then we solve the equations of motion to find the optimal path in terms of $f$ and $\alpha$. We put them back in the complexity functional to find the circuit complexity of the optimized path from reference to the target state in the manifold of unitary transformations. First, we write the complexity functional in terms of $f$ instead of the inverse diffeomorphism $F$.
\bea{comfuncf}
\non \mathcal{C}[f,\a]&=&\frac{1}{2\pi}\int d\s dt \Biggl[ \left(\frac{\dot{f}}{f^\prime} \right) \bigg\{ \biggl( j_0(\s)+\frac{c_1}{48\pi}\left(\frac{f^{\prime\prime}}{f^\prime}\right)^\prime \biggr)+ \biggl( 2 \a^\prime (\s) p_0(\s)+2p_0^\prime\a(\s)-\frac{c_2}{24\pi}\a^{\prime\prime\prime}(\s)\biggr)\bigg\}\\
&&~~~+\frac{1}{f^{\prime^2}}\left(\dot{\a}(\s)+\a(\s)\dot{f}^\prime \right) \left(p_0(\s)+\frac{c_2}{24\pi}\{f,\s \}\right) \Biggr] .
\eea
Next, we extremize wrt $\a$ and obtain the equation of motion for $\a(t,\s)$.
\be{}
\frac{\p \mathcal{C}}{\p \a}-\frac{\p}{\p t}\left(\frac{\p \mathcal{C}}{\p \dot{\a}} \right)-\frac{\p}{\p \s}\left(\frac{\p \mathcal{C}}{\p {\a}^\prime} \right)+\frac{\p^2}{\p \s^2}\left(\frac{\p \mathcal{C}}{\p {\a}^{\prime \prime}}\right)-\frac{\p^3}{\p \s^3}\left(\frac{\p \mathcal{C}}{\p {\a}^{\prime \prime \prime}}\right)=0\,.
\ee
 Such that,
 \be{eom1}
2 p_0^\prime \frac{\dot{f}}{f^\prime}+\frac{\dot{f}^\prime}{f^{\prime^2}}\left(p_0+\frac{c_2}{24\pi}\{f,\s\} \right)- \frac{\p}{\p t}\bigg\{\frac{1}{f^{\prime^2}}\left(p_0+\frac{c_2}{24\pi}\{f,\s\} \right) \bigg\}-2\biggl(p_0 \;\frac{\dot{f}}{{f}^\prime}\biggr)^\prime+\frac{c_2}{24\pi}\biggl(\frac{\dot{f}}{f^\prime}\biggr)^{\prime\prime\prime}=0\,.
 \ee
Without the loss of generality, we assume
 \be{}
 j_0=\textrm{const},\quad p_0=\textrm{const}.
 \ee
Then the simplest solution comes as,
 \be{}
 \frac{\dot{f}}{f^\prime}=\textrm{const}^{\footnotemark}, ~~\dot{f}^\pr=0,~~f^{\pr}\neq0,~~f^{\pr\pr}=0\,.
 \ee
 Thus,
 \be{}
 f(\s,t)=a_0 \s+a_1 t\,.
 \ee
 Also, periodicity demands that
 \bea{fsol}
\non  f(\s+2\pi)&=&f(\s)+2\pi,~~\implies a_0=1\\
\text{Thus,~} f(\s,t)&=&\s+a_1 t\,. 
 \eea 
 \footnotetext{It is to note here, that similar solutions are also obtained for the case of Virasoro \cite{Caputa:2018kdj}(before adding the contribution $c\beta(t)$ due to the central term) and Warped CFT \cite{Bhattacharyya:2022ren}.}
 Also, extremizing wrt $f(t,\s)$, we get the other equation of motion,
 \be{}
 \frac{\p \mathcal{C}}{\p f}-\frac{\p}{\p t}\left(\frac{\p \mathcal{C}}{\p \dot{f}} \right)-\frac{\p}{\p \s}\left(\frac{\p \mathcal{C}}{\p {f}^\prime} \right)+\frac{\p^2}{\p t \p \s}\left(\frac{\p \mathcal{C}}{\p \dot{f}^{ ^\prime}} \right)
 +\frac{\p^2}{\p \s^2}\left(\frac{\p \mathcal{C}}{\p {f}^{\prime \prime}}\right)-\frac{\p^3}{\p \s^3}\left(\frac{\p \mathcal{C}}{\p {f}^{\prime \prime \prime}}\right)=0\,.
 \ee
 \bea{fsol2}
\non &&\frac{2}{f^\prime}\Biggl[j_0\left(\frac{\dot{f}}{f^\prime} \right)^\prime -\frac{c_1}{48\pi} \left(\frac{\dot{f}}{f^\prime} \right)^{\prime\prime\prime}\Biggr]+\frac{1}{f^\prime}\Bigg( 2\dot{f}^{ ^\prime}-\frac{\dot{f}}{f^\prime}\biggl\{\frac{2f^{\prime\prime}}{f^\prime}-\p_\s \biggr\}+\p_t \Biggr)\Bigg( 2\a^\prime\: p_0+2p_0^\prime \:\a-\frac{c_2}{24\pi}\a^{\prime\prime\prime}\Biggr)\\
\non && +\frac{\p}{\p \s}\Biggl[\frac{1}{f^{\prime^2}} (\dot{\a}+\a \dot{f}^{ ^\prime})\biggl\{ \frac{2}{f^\prime} (p_0+\frac{c_2}{24\pi}\{f,\s\})-\frac{c_2}{24\pi}\left(-\frac{f^{\prime\prime\prime}f^{\pr\pr}}{f^{\pr^2}}+3\frac{f^{\pr\pr^3}}{f^{\pr^3}} \right)
\biggr\} \Biggr]\\
\non&&-\frac{\p^2}{\p t \p \s}\Biggl\{-\frac{\a}{f^{\pr^2}}(p_0+\frac{c_2}{24\pi}\{f,\s\}) \Biggr\}-\frac{c_2}{24\pi}\Biggl[\frac{\p^2}{\p \s^2}\Biggl\{3 \frac{f^{\pr\pr}}{f^{\pr^3}} (\dot{\a}+\a \dot{f}^{ ^\prime}) \Biggr\} -\frac{\p^3}{\p \s^3}\Biggl\{ \frac{1}{f^{\pr^3}} (\dot{\a}+\a \dot{f}^{ ^\prime}) \Biggr\} \Biggr]=0\,.\\
 \eea
The first term in parentheses is the equation of motion for the diffeomorphism in the case of Virasoro circuit complexity \cite{Erdmenger:2020sup}. This \eqref{fsol2} is a highly non-linear and coupled PDE and we write only the simplest solution, 
 \be{}
 \dot{\a}^\pr=0,~~\a^{\pr\pr}=0~~\implies \a(\s,t)=a_2 \s+a_3(t).
 \ee
 Imposing the periodicity condition,
 \be{alphasol}
  \a(\s+2\pi)=\a(\s),~~\implies a_2=0, \text{thus,~}\a=a_3(t).
 \ee
The constant $a_1$ and the function $a_3(t)$ remain undetermined. Interestingly, the simplest solutions found in \eqref{fsol} and \eqref{alphasol} agree with the solutions to gravitational saddle points with constant  orbit representatives $j_0,p_0$ in \cite{Merbis:2019wgk}.  For flat space cosmologies (FSC), the solution is given by 
\be{}
f=\sigma,~~\a=t.
\ee

\smallskip
\noindent
The FSC solutions, together with its BMS$_3$ descendents, correspond to a generic (massive) coadjoint orbits of the BMS$_3$ group, in the same way that the BTZ solutions correspond to the Virasoro coadjoint orbits. For the FSC solutions, we have the orbit representatives $p_0>0$ and $j_0\neq 0$. We only consider evolution inside such orbits, consisting of descendants of the initial orbit representatives, obtained by acting with BMS generators. \footnote{Field theory states representing FSC solutions cannot be obtained from a generic primary, but only from a primary that already has the correct expectation values for the zero modes of the BMS algebra $j_0,p_0$ \cite{Merbis:2019wgk}.}

\medskip\noindent
Putting the solution back to the complexity functional we obtain, the circuit complexity for the optimal path
\bea{}
\mathcal{C}[T]&=&\frac{1}{2\pi}\int_{0}^{2\pi}\: d\s\: \int_{0}^{T} dt \big[ j_0 a_1+\dot{a_3}p_0\big]\\ 
\non&&  \\
\label{compans}
&=& a_1\; j_0 T+p_0(a_3(T)-a_3(0)).
\eea
  Here, $j_0=|\D-\frac{c_1}{24\pi}|, \:\: p_0=|\xi-\frac{c_2}{24\pi}|$, the expectation values of $L_0$ and $M_0$ on the cylinder. $\D$ and $\xi$ are the eigenvalues of $L_0$ and $M_0$ respectively. Similar to the Virasoro (as well as Warped CFT) case, the BMS complexity \eqref{compans} depends on the central charges and is proportional to $T$. For FSC, $a_1=0$ and $a_3(t)=t$, hence,
  \be{comres}
  \mathcal{C}[T] \propto\; p_0 \; T.
  \ee

  
\section{Conclusions}\label{discussion}
We have computed Nielsen's complexity for a non-Lorentzian field theory in this paper. We have considered $2d$ Carrollian CFT (BMS$_3$ invariant field theory), a putative dual to $3d$ asymptotically flat spacetimes. We have taken two distinct approaches here, namely limiting and intrinsic. For the relativistic cousin of the said field theories, the circuit complexity functional takes the form of the geometric action on the co-adjoint orbit of the Virasoro group. Using the limiting approach on the relativistic CFT complexity functional, we show that it is possible to write the geometric action on the co-adjoint orbit of the BMS$_3$ group as its complexity functional for BMS circuit complexity. However, this does not give us the full picture. \footnote{The notion of ``complexity=geometric action" holds for BMS$_3$ only through the limiting analysis, and not in the intrinsic approach. This result, for a semi-direct product group like BMS$_3$, is unlike the ``complexity=geometric action" proposal for direct product groups (e.g Virasoro and Kac-Moody \cite{Erdmenger:2020sup}).}Hence, we refine our analysis using the intrinsic approach. We consider the BMS symmetry transformations as the required gates and find the instantaneous velocities from the BMS group product law. As a result, the complexity functional acquires a more involved structure than only the geometric action for BMS. We show that even after the addition of the Hamiltonian to the BMS geometric action, it is not possible to reproduce the exact form of the complexity functional. However, for a particular value of the instantaneous velocity $\e_M(\t,\s)$, we can relate the complexity functional with available literature (modified bulk Chern-Simons classical gravity action for three-dimensional flat spacetime reduced with BMS boundary conditions to a two-dimensional boundary theory). Next, we extremise the complexity functional with the group paths and find the equations of motion to be highly non-linear and coupled PDEs. We write the simplest solutions and find the similarity with the solutions to gravitational saddle points for flat space cosmology. This provides a consistency check for our computations and connects with (flat-space) holography. It will be interesting to investigate whether there are any more non-trivial solutions of (\ref{fsol2}). Also, the result in equation \eqref{comres} shows that the BMS circuit complexity (calculated using the symmetry generators as the gates) grows linearly in time. This reflects the similar behaviour for the Virasoro complexity in \cite{Caputa:2018kdj}.  This also supports the general claim arising from holography about the linear growth of the complexity. The holographic complexity proposal typically suggests a volume divergence. But this way of computing the complexity does not give rise to the divergence, which is similar to the earlier results from the findings in \cite{Caputa:2018kdj,Erdmenger:2020sup, Bhattacharyya:2022ren} for Virasoro and warped CFT. This absence of the divergence structure needs to be understood better. It will be interesting to find the exact reason behind this in future.

\paragraph{Discussion and future directions}

\paragraph{Complexity and Berry Phases:} In \cite{Caputa:2018kdj,Erdmenger:2020sup}, it was shown that the complexity functional for the Virasoro group and the complexity for all possible diffeomorphisms $f(t,\s)$ of the Virasoro group is equivalent to a Berry phase upto a boundary term.\footnote{The authors also mentioned that not all Berry Phases are equivalent to complexity, particularly the non-optimal closed paths $f(t,\s)$, which do not solve the equations of motion for the Virasoro case.} We briefly comment on the relation between BMS Berry Phases and complexity in parallel to the relativistic case. The BMS Berry phase is given by \cite{Oblak:2017ptc}
\bea{}
\mathcal{B}_{\text{scalar}}[f,\a]&=&-\frac{M}{2\pi}\int \:dt\: d\s\: \frac{\dot{\a}(t,f(t,\s))}{f^\pr}-\frac{c_2}{24\pi}\int \: dt\: d\s\: \frac{\dot{\a}(t,f(t,\s))}{f^\pr}\left(\{f,\s\}-\frac{1}{2} \right).\\
\mathcal{B}_{\text{spin}}[f]&=&-\frac{1}{2\pi}\int \:dt\: d\s\: \frac{\dot{f}}{f^\pr}\left[s-\frac{c_1}{24}+\frac{c_1}{24}\left( \frac{f^{\pr\pr}}{f^\pr}\right)^{\pr} \:\right].
\eea
Even though the spin part of the BMS Berry phase is seen in the complexity functional \eqref{comfuncf}, the scalar part is not apparent due to the change of argument of $\dot{\a}$. Also, we obtain an extra part (the second term in parentheses of the complexity expression \eqref{comfuncf}) that is unclear in the context of the BMS Berry Phase. We leave this for future investigation.

\paragraph{Higher dimensional exploration:} It would be interesting to look into aspects of circuit complexity for $d>2$ BMS invariant field theories. Recently, in \cite{2022JHEP...12..154B} the geometric action for BMS$_4$ was given. A natural extension would be to investigate whether any similarity exists for the higher dimensional geometric action and complexity functional.

\paragraph{Other applications:}  There are other notions of complexity based of various other methods, e.g. Fubini Study distance, Path Integral \cite{Erdmenger:2021wzc,Flory:2020eot,Caputa:2017yrh,Bhattacharyya:2018wym,Camargo:2019isp}. It will be very interesting to generalize these notions (particularly the Fubini-Study one) of complexity for BMS invariant field theory. Furthermore, in recent times, the notion of `Krylov Complexity' has gained much attraction. It is related to the notion of the complexity associated with the operator growth \cite{2019PhRvX...9d1017P}. It has been recently studied for CFT \cite{Dymarsky:2021bjq,2022arXiv221214429A,2022arXiv221214702C} and for some quantum mechanical models where Carroll symmetry emerges \cite{Banerjee:2022ime}. It would be exciting to explore different notions of complexities in other areas, e.g. Carrollian physics in the context of flat bands \cite{Bagchi:2022eui}, or with the inclusion of Super Carrollian structures \cite{Bagchi:2022owq}. It will also be of importance to find other optimal solutions of the equations of motion \eqref{eom1}, \eqref{fsol2} for the group paths $f,\a$ for Carrollian circuit complexity.  We hope to make a systematic of these for BMS field theories in the near future.

\section*{Acknowledgements}

A.B thanks Shubho Roy and Aritra Banerjee for useful discussions and collaboration on related projects. PN is thankful to Daniel Grumiller and Max Reigler for their helpful comments. Research of A.B and PN is supported by the Relevant Research Project grant (202011BRE03RP06633-BRNS) by the Board Of Research In Nuclear Sciences (BRNS), Department of Atomic Energy (DAE), India. Research of A.B is also supported by the Mathematical Research Impact Centric Support Grant (MTR/2021/000490) by the Department of Science and Technology Science and Engineering Research Board (India). PN is also indebted to the Science and Engineering Research Board (SERB-India) for the International Travel Support Grant (ITS/2022/000812) during the progress of this work, and also to the people of India for their generous support to research in basic sciences.


\appendix

\bibliographystyle{JHEP}
\bibliography{bmscomp}

\end{document}